\documentclass[twocolumn,aps,showpacs,prb]{revtex4}
\usepackage{epsfig,amsmath}

\renewcommand{\section}[1]{\noindent{\bf #1 -- }}
\renewcommand{\subsection}[1]{\noindent{\sl #1 -- }}

\def\be{\begin{equation}} 
\def\ee{\end{equation}} 

\newcommand{\pf}{P_{0,{\rm f}}}

\begin{document}

\title{Particle tunneling through a polarizable insulator}

\author{Peter Nalbach$^1$}
\author{Walter Harrison$^2$}
\affiliation{$^1$Department of Physics, $^2$Department of Applied Physics,
  Stanford University, Stanford, California 94305, USA} 
\date{\today}
\begin{abstract}
The tunneling probability between two leads
connected by a molecule, a chain, a film, or a bulk polarizable insulator
is investigated within a model of an electron tunneling from lead A to
a state higher in energy, describing the barrier, and from there to lead B. To
describe the possibility of energy exchange with excitations of the molecule
or the insulator we couple the intermediate state to a single oscillator or to
a spectrum of these, respectively. In the single-oscillator case we find for
weak coupling that the tunneling is weakly suppressed by a Debye-Waller-type
factor. For stronger coupling the oscillator gets {\it stiff} and we 
observe a suppression of tunneling since the effective barrier is
increased. The probability for the electron to excite the oscillator increases
with the coupling. In the case of a film, or a bulk barrier the behavior is
qualitatively the same as in the single oscillator case. 
An insulating chain, as opposed to a film or a bulk connecting the two leads,
shows an {\it orthogonality catastrophe} similar to that of an electronic
transition in a Fermi gas.
\end{abstract}

\maketitle

There is a vast interest in the electronic properties of molecular systems or
tunnel junctions due to their potential use for electronic
applications. Thereby, a tunnel junction might be a microscopic or macroscopic
one-, two- or three dimensional insulator connecting two metal leads. As long
as the molecule or the tunnel junction is not excited by the electron the
conductance can be phenomenologically described by introducing a matrix
element for tunneling through the barrier. However, the possibility to excite
the barrier opens new paths for the electron, thus, enhancing the
conductance. In recent experiment on molecular junctions \cite{Bao02,Park00} 
the conductance shows several steps with increasing bias voltage between the
two leads underlining the importance of vibrations excited by the tunneling
electron \cite{Schoel01}. 
We are solely interested in the influence of vibrations on the conductance
and we do not restrict ourselves to weak coupling between the vibrations
and the tunneling electron.
In detail we investigate the probability for a tunneling electron to excite 
the molecule/tunnel junction. Therefore we introduce an intermediate state in
the junction which itself is coupled to either a single oscillator to mimic the
excitations of a molecule or to a spectrum of oscillators to mimic photonic
excitations of a tunnel junction. A similar approach allowing two paths to a
final state for a tunneling particle was used by Harrison \cite{Walter1994} to
investigate how the coupling destroys the coherence between two paths. 

\section{A Single Oscillator} 
The electron starts in lead A from an initial state $|0\rangle$ with
energy $E_0$. 
This state is coupled to the intermediate electron state $|1\rangle$ with
energy $E_1\ge E_0$ by the tunneling matrix element $V_{01}$.  
The intermediate state is also coupled by a matrix element
$V_{1\alpha}/\sqrt{M}$ to $M$ final states $|\alpha\rangle$ with $\alpha\ge 2$
in lead B. 
For sufficiently small tunneling matrix elements the transition probability
$\pf$ for the electron tunneling from $|0\rangle$ through $|1\rangle$ to
a final state $|\alpha\rangle$ can be obtained by simple second-order
time-dependent perturbation theory \cite{LandauLifshitz} as
\be\label{eq1}
 \pf=\frac{2\pi}{\hbar} \frac{1}{M}\sum_\alpha \left| \frac{
 V_{1\alpha}V_{01}}{E_1-E_0} \right|^2 \,
\delta(E_0-E_{i}) = \frac{2\pi V^2_{1\alpha}V_{01}^2}{b\hbar\epsilon^2} 
\ee
with $\epsilon=E_1-E_0\gg V_{01}, V_{1\alpha}$. For the summation over all
final electron states we assumed a uniform density
of states of the $M$ states within a band of bandwidth $b$ in lead B. 

An interaction $H_I$ between the intermediate state and the excitations of
a molecule can be mimicked by coupling the state $|1\rangle$ to a single
harmonic oscillator: $H_I=\gamma |1\rangle\langle 1| \hat{x}$ where $\gamma$ is
a coupling constant and $\hat{x}$ is the displacement coordinate of the
oscillator. The quantum mechanical oscillator states when the electron is in
one of the leads are denoted as $|\phi_i\rangle$, where $i$ is the excitation
number, whereas $|\psi_i\rangle$ is the 
oscillator state with $i$ excitations when the electron is in the intermediate
state. 
The total state of the problem can be expressed by the states
$|0,\phi_i\rangle$, $|1,\psi_i\rangle$ and $|2,\phi_i\rangle$ with energy
$E_{\nu=0,1,\alpha}+i\hbar\omega$ (the zero-point energy is absorbed in the
$E_\nu$). 

We consider only initial states without excitations. The probability
$P_{00,{\rm f}n}$ for tunneling from state $|0\rangle$ with zero excitations to
a final state with $n$ excitations in the oscillator is modified from
Eq. (\ref{eq1}) as
\be\label{eq2} P_{00,{\rm f}n} \,=\, \pf\cdot
\left| \sum_{i} \frac{\epsilon}{\epsilon+ i \hbar\omega}
\langle \phi_n|\psi_i \rangle\langle\psi_i|\phi_0\rangle \right|^2 \; ,
\ee
where we sum over all possible numbers $i$ of virtual excitations in
the oscillator in the intermediate state.

The oscillator states $\psi_i(x)=\phi_i(x-x_0)$ are shifted by
\be\label{eq6} x_0\,=\, \frac{\gamma}{m\omega^2} \ee
since an electron in the intermediate state shifts the potential minimum
position of the oscillator. With the zero point fluctuation
\be\label{eq8} a_0 \,=\, \sqrt{\frac{\hbar}{m\omega}} \ee
of the harmonic oscillator we obtain for the overlaps
\begin{eqnarray} \label{eq7}
&& \langle \phi_n|\psi_i \rangle \,=\,
 \exp\left(-\frac{x_0^2}{4a_0^2}\right) \,\cdot \\
&& \sum^n_{j=\left\{ \mbox{\tiny $\begin{array}{cc} 0 & i>n \\
      n-i & i<n \end{array}$\normalsize} \right. } 
\frac{(-1)^j \sqrt{\frac{i! n!}{2^{i-n+2j}}} }{j! (i-n+j)!  (n-j)!}   
\left( \frac{x_0}{a_0} \right)^{i-n+2j} \; . \nonumber
\end{eqnarray}
Assuming that the oscillator is in the ground state before and after
tunneling the expressions simplify
substantially and we obtain for the tunneling probability
\be\label{eq9} P_{00,{\rm f}0}=\pf\cdot
\left| \sum_{i} \frac{1}{i!} W^{i}
  \frac{\epsilon}{\epsilon+ i \hbar\omega} \right|^2 
\exp\left(-2W\right) \; ,
\ee
defining
\be\label{eq11} W \,=\, \frac{1}{2} \left(\frac{x_0}{a_0}\right)^2 \; .\ee
The sum runs over the number $i$ of virtual excitations in the intermediate
state. The contributions peak for $i\simeq W$ virtual excitations. 
In the weak-coupling limit, $W\ll 1$, tunneling without virtual excitations,
$i=0$, dominates and the tunneling is influenced by a Debye-Waller type term
$\sim\exp(-2W)$. 
For strong coupling, $W\gg 1$, the contribution to the sum peaks at $i\sim W$
and noting $\sum_iW^i/i!\sim e^W$ we obtain 
\be\label{eq12} P_{00,{\rm f}0} \simeq \pf\cdot
\left| \frac{\epsilon}{\epsilon+ W \hbar\omega} \right|^2 =
\frac{2\pi}{\hbar b}\frac{V_{1\alpha}^2V_{01}^2}{(\epsilon+W\hbar\omega)^2} 
\ee
which numerically turns out to be very precise when $W\gg 1$.
The tunneling electron
leaves the oscillator unmoved but its potential minimum position is
shifted putting the oscillator in an excited state with an excitation energy
$\frac{1}{2}m\omega^2x_0^2=W\hbar\omega$. 
Accordingly, the barrier is
effectively raised by this energy.

Naturally, we would assume that the crossover at $W\sim 1$ coincides
with the region where $P_{00,{\rm f}0}$ falls from its value without coupling
to an oscillator to a strongly suppressed probability. 
However, Eq. (\ref{eq12}) shows that even for $W\gg 1$ the coupling to
the oscillator only influences the tunneling probability substantially for
$W\hbar\omega>\epsilon$. 
Our perturbative approach limits us to energies smaller than the barrier
heights. Therefore we cannot investigate the strong coupling limit
$W\hbar\omega\ge\epsilon$ and we are confined to oscillator frequencies
$\omega<\epsilon/\hbar$. 

Defining a number $n_r$ with
\be\label{eq10} n_r\hbar\omega \,=\, \epsilon \ee
we find that $P_{00,{\rm f}0}$ is a function of $W/n_r$ instead of $W$ and we
are restricted to the effective weak coupling limit $W/n_r\le 1$.
 
If we interpret the virtual excitation energy $W\hbar\omega$ as the energy put
into the oscillator during tunneling we expect that in the resonant case when
$W=1$ the probability for exciting the oscillator will be maximal. 
However, we saw that for $\hbar\omega\le\epsilon$ the influence on the
electron from the coupling to the oscillator increases with $W/n_r$. We
numerically find that the probability $P_{00,21}$ for exciting the
oscillator with a single excitation increases linear with $W/n_r$ (quadratic
with the coupling $\gamma$ between electron and oscillator) and
the probability $P_{00,2n}$ decreases for a higher number of excitations $n$. 
If we expand
$\epsilon/(\epsilon+i\hbar\omega)\approx 1-i\hbar\omega/\epsilon$ in
Eq. (\ref{eq2}) up to first order in $\hbar\omega/\epsilon$ we obtain
\be \label{eq101} P_{00,{\rm f}n} \simeq \pf\cdot
\left| \delta_{n0}\, + \langle\phi_n| \frac{\gamma \hat{x}}{\epsilon}|
  \phi_0\rangle  \right|^2 
\ee
which shows the quadratic behavior with the coupling $\gamma$. Summing over
all possible numbers $n$ of excitations we get for the probability to excite
the oscillator $P_{00,{\rm f}n\ge1}\approx(\gamma a_0/\epsilon)^2\pf $.
This approximation has the same limit as discussed before namely that the
virtual energy $W\hbar\omega$ of the mainly contributing tunneling path is
smaller than the barrier heights $\epsilon\ge W\hbar\omega$. Therefore this
result holds only for couplings where 
$(\gamma a_0/\epsilon)^2\le(2\hbar\omega/\epsilon)\pf$. 

\section{Multiple Oscillators}
\begin{figure}[t]
\vspace*{5mm}\epsfig{file=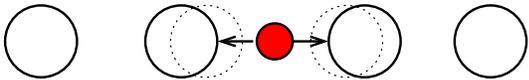,width=7cm}
\caption{\label{fig11} A linear chain of atoms is illustrated in which a
  tunneling electron might hop to an anti-bond between atom $m$ and $m+1$
  repelling both neighbors.}  
\end{figure}

In the next step we introduce a spectrum of oscillators coupled to the
tunneling electron to investigate the tunneling of an electron through a
tunnel junction, for example a linear chain of atoms as illustrated in
Fig. \ref{fig11}. 
Assuming that all bonding states between the atoms of the chain are occupied
by electrons, the tunneling electron hops to an anti-bonding state between two
atoms. The electron repels with force $F$ the nearest neighbors resulting in a
coupling  
\[ H_I \,=\, - |1\rangle\langle 1|\cdot F \cdot(\delta x_{m+1}-\delta x_m) \]
of an electron in state $|1\rangle$ to the displacement $\delta
x_{m+1}$ and $\delta x_m$ of the nearest neighbors.
For a chain of $N$ atoms and lattice constant $d$ we obtain, using fixed
boundary conditions at position $r=0$ and $r=(N+1)d$, $N$ phonon modes with
$\delta x_m=\sqrt{2/N}\sum_ku_k\sin(kx_m)$ and 
\be\label{eq30} k \,=\, k_n \,=\, \frac{n}{N+1} \,\frac{\pi}{d}
\quad\mbox{with}\quad n=1..N  \; .
\ee
It is convenient to take $N$ even and the anti-bond $m=N/2$.
Then the coupling between the electron and the phonons turns out to be 
\[ H_I = |1\rangle\langle 1|\cdot \sum_{k_n} \lambda_{k_n}u_{k_n} \]
with the coupling constants
\be\label{eq14} \lambda_{k_n} \,=\, \left\{ \begin{array}{ll}
    2F\cdot \sqrt{\frac{2}{N}} \sin\left(\frac{k_nd}{2}\right) &
    :n\;\mbox{odd} \\ 0 & :n\;\mbox{even} \end{array} \right. \; .
\ee
In the following we use the indices $k_n$, $k$ and $n$ interchangeably. The 
coupling constant $\lambda_n$ between the
electron and the mode with wave vector $k_n$ is proportional to
$1/\sqrt{N}$, so in the thermodynamic limit ($N\longrightarrow\infty$) the
coupling effects estimated by $\sum_n\lambda_n^2$ stay finite. 

In the case of a single oscillator the behavior of the system was
determined by $W=\frac{1}{2}(x_0/a_0)^2$. Since $\lambda_n\sim 1/\sqrt{N}$ in
the case of many phonons each
single mode is weakly coupled to the electron for large $N$. However, the
influence to the tunneling electron is now determined by the sum over all modes
$W=\frac{1}{2}\sum_n(x_n/a_n)^2$ with $x_n=\lambda_{k_n}/(m\omega_n^2)$ and
$a_n=\sqrt{\hbar/(m\omega_n)}$. The equations given above are for a
one-dimensional chain but they are easy to 
generalize to two and three dimensions. It is a little intricate to work out
the coupling but essentially the coupling constants $\lambda_k$ look the same
and only the density of phonon modes $g(k)$ depends on the dimensionality,
$g(k)\propto k^{d-1}$. This turns out to give a striking difference between
the one dimensional and the other cases. 

With no phonon in the initial state $\{m_k=0\}$ we find for the transition
probability to the final state with $l_k$ excitations in mode $k_n$
\begin{eqnarray}\label{eq25}P_{0\{m_k=0\},{\rm f}\{l_k\}} 
\,=\, \pf\cdot\hspace*{3.5cm} \\ 
\left| \sum_{\{i_k\}}
\frac{\epsilon}{\epsilon+\sum_ki_k\hbar\omega_k}\, 
\prod_k \langle \phi_{l_k}(k)|\psi_{i_k}(k)\rangle\langle
\psi_{i_k}(k)|\phi_0(k)\rangle  \right|^2 \nonumber
\end{eqnarray}
where $\phi_{l_k}(k)$ and $\psi_{l_k}(k)$ are the wavefunctions of the
unshifted and shifted mode $k$ with $l$ excitations respectively.
The sum $\sum_{\{i_{k}\}}$ sums over all possible total transient
states with virtual excitation numbers $i_{k}$ for each mode. 
For each of these states the product $\prod_{k}$ over the various
overlaps calculates the total overlap between the initial and the transient
state times the overlap of the transient and the final state.

Let us focus first on the probability $P_{0\{m_k=0\},{\rm f}\{l_k=0\}}$ for
tunneling without final state excitations. 
The contribution $A_0$ from the tunneling
'path' without virtually exciting any phonon is given by the overlap between
the ground states of the unshifted and shifted oscillator modes
\[ A_0= \prod_k \langle \phi_0(k)| \psi_0(k) \rangle \langle \psi_0(k)|
\phi_0(k) \rangle \,=\, e^{-\frac{1}{2}\sum_n\frac{x_n^2}{a_n^2}}\; . 
\]

\subsection{Einstein Approximation}
An Einstein approximation leaves us essentially with $N$ modes with the same
frequency $\omega_E$ and coupling constant $\gamma$.
We find 
\be\label{eq71} W \,=\, \frac{1}{2}\sum_n\frac{x_n^2}{a_n^2} \,=\,
\frac{1}{2}\frac{\gamma^2}{\hbar m \omega_E^3} \; .
\ee
which is identical to the single oscillator result.
For the contributions $A_m$ from tunneling paths involving $m$ virtual
excitations we obtain 
%
\begin{equation}\label{eq99}
A_m = W^m \frac{1}{m!} e^{-W} \cdot \frac{\epsilon}{\epsilon+m\hbar\omega_D}
\; .
\end{equation}
The factorial $1/m!$ results from the fact that, once we chose $m$ oscillators
to put the $m$ virtual excitation in, there are still $m!$ ways of
distributing the virtual excitation to the oscillators. Physically, however all
these $m!$ ways are identical. Technically it is convenient to sum over all
these and we obtain $m!A_m$ which we then divide by the factorial to get
Eq. (\ref{eq99}). Note also that $A_m$ includes only tunneling path where all
$m$ excitations are in $m$ different modes. Tunneling paths with more than
one excitation in a mode are suppressed for large $N$ by factors $1/N$. 
The probability $P_{0\{m_k=0\},{\rm f}\{l_k=0\}}$ for tunneling without
final state excitations thus is equivalent to Eq. (\ref{eq9}). 
An Einstein spectrum yields the same tunneling probability as a single
oscillator if no mode is excited.  

The probability for tunneling through the chain and exciting mode $j$ with a
single excitation is given by 
\be\label{eq20} P_{00,{\rm f}1_j} \,=\, \left(\frac{x_j}{a_j}\right)^2 \,
e^{-\frac{x^2_j}{2a^2_j}}\; P'_{00,20} \; .
\ee
$P'_{00,20}$ is the probability for the electron tunneling through the
chain without final state excitations calculated without including mode $j$. 
Since a single mode does not change results appreciably we
take $P'_{00,20}\simeq P_{00,20}$. 
We sum $j$ over all modes. The exponential is unity for large $N$ and we use
Eq. (\ref{eq71}) to find for the probability to excite $n$ modes
\be\label{eq81} P_{00,{\rm f}n} \,=\, \frac{1}{n!} W^n \; P_{00,20} \; .
\ee
Note that contributions from tunneling path with more than one excitation in a
single mode are suppressed. 
The factorial $n!$ has the same origin as $m!$ in Eq. (\ref{eq99}).
For $W<1$ tunneling through the chain without exciting it and without using
virtual excitations dominates. However, for $W>1$ the tunneling electron uses
predominantly $W$ virtual excitations to tunnel and the chain has most likely
$W$ excitations once the electron tunneled through. 
This shows a difference from the single oscillator case where in
the strongly coupled regime, $W>1$, the oscillator was not excited by
the tunneling electron. Only for $W/n_r$ near one excitations are
likely. To understand this we note that each single mode in the
many oscillator case is weakly coupled. The probability to excite some is only
finite because there are many modes.

\subsection{Debye Approximation}
Within a Debye approximation we assume a spectrum of modes with linear
dispersion $\omega=vk$ with the speed of sound $v$.
The Debye cut-off $k_D=(N/N+1)(\pi/d)$ is chosen to give the correct number
of modes. 
It leads qualitatively to the same results as an Einstein
approximation for a two and a three dimensional tunnel junction since their
phonon spectra favor high frequency modes and $\omega_E$ is replaced by an
average frequency. We therefore focus on tunneling
through a chain of atoms. We further restrict the following considerations to
the probability for tunneling without final state excitations since the
probability for tunneling with final state excitations follows then from
Eq. (\ref{eq81}).
   
The chain has a constant density of states and we find
\be W = \frac{1}{2}\sum_n\frac{x_n^2}{a_n^2} = \frac{k_c}{k_D}\,
\sum_{j=1}^{N/2} \frac{1}{j} \sim \frac{k_c}{k_D}\, \ln(N)
\ee
with $k_c=(Fd)^2/(\hbar m v^3)$.
For large $N$ tunneling through the chain without the help of virtual
excitations is  
suppressed by $\exp(-2W)\sim\exp(-2(k_c/k_D)\ln(N))=N^{-(2k_c/k_D)}$. 
In the thermodynamic limit, $N\rightarrow\infty$, this approaches zero. 
This behavior is
similar to the 'orthogonality catastrophe' described by Anderson
\cite{PWA} for an electron tunneling into a Fermi gas which has a
constant density of states near the Fermi surface.  

We now investigate the contribution $A_m$ from a tunneling path involving $m$
virtual excitations. 
The key question is whether there is a
dominant tunneling path (a $m$ with $A_m$ maximal) and what would be its
effective virtual energy $\sum_ki_k\hbar\omega_k$ (compare Eq. (\ref{eq25})). 
We suggest that the contribution of mode $j$ to the effective
virtual energy is determined by a factor $(x_j^2/a_j^2)$ since the same factor
determined in the case of tunneling with one final state excitation the
contribution of the tunneling path with mode $j$ finally excited (compare
Eq. (\ref{eq20})). Since $A_m$ involves $m$
virtual excitations we introduce further a 'normalizing' factor $m/W$ and find 
\be\label{eq21} A_m \,=\,  e^{-W} W^m \frac{1}{m!}
\frac{\epsilon}{\epsilon+\frac{1}{2}\sum_j\frac{mx_j^2}{Wa_j^2}\hbar\omega_j}
\; . 
\ee
For the energy of the $m$ virtual excitations we find
\be\label{eq78} 
    \sum_j\frac{mx_j^2}{Wa_j^2}\,\hbar\omega_j \,=\, \frac{m}{W}\hbar\omega_c
    \; . 
\ee
Numerically we find that this approximation holds when the energy of the $m$
virtual excitations does not exceed the barrier heights. 
For large finite $N$ we find
that $W$ tends logarithmically to infinity. However, for $W\gg 1$ the
probability for tunneling without final state excitations is simply given
by the tunneling probability through an effective barrier (compare Eq. 
(\ref{eq12})) enhanced by the energy of the virtual excitations of the
dominating tunneling path. The dominating path involves $W$ virtual
excitations leading to an energy $\hbar\omega_c$ of the virtual excitations
which is independent of $W$. Thus the probability for tunneling
through the chain without final state excitations is finite and independent of
the length $N$. 

Note that the maximal virtual energy $W\hbar\omega_D$ possibly used in a
tunneling event with $W$ virtual excitations increases logarithmically with
the number $N$ of modes. Once this energy $W\hbar\omega_D$ exceeds the barrier
heights $\epsilon$ the numerical investigation showed that our approximation,
Eq. (\ref{eq21}), fails. However at the same point our approach using second
order perturbation theory is questionable. We therefore believe that our
approximative result holds in the thermodynamic limit and that tunneling
through a chain without final state excitations is possible. 

In conclusion, we first investigated tunneling through a barrier with a
coupling to a single oscillator. The result can be characterized by a coupling
strength $W$ given by the square of the ratio of the classical
shift of the oscillator to its zero-point fluctuation. For weak coupling,
$W<1$, tunneling without final state excitations is suppressed by a factor
$e^{-2W}\approx 1-2W$. For strong coupling, $W>1$, tunneling is
suppressed by an increase in the tunneling barrier height by an energy $W$
times the vibrational energy $\hbar\omega$. The probability for tunneling with
final state excitations increases in proportion to $W$.
We then introduced a barrier with coupling to a spectrum of $N$ modes, with
couplings proportional to $1/\sqrt{N}$. It is shown that in an Einstein
approximation, where all modes have the same  frequency $\omega_E$, the
results are essentially the same. The same holds true for two or three
dimensional Debye spectra where the frequency is proportional to the wave
number. However, a one-dimensional chain shows an {\it orthogonality
  catastrophe} similar to that of an electronic transition in a Fermi gas
since $W\propto\ln(N)$. The probability for
tunneling without final state excitations and without the help of
virtual excitations is suppressed by a factor $1/N$ raised to a finite power. 
However, our investigation 
suggests that the probability for tunneling without final state excitations
but with the help of virtual excitations stays finite for $N\rightarrow\infty$.

P. Nalbach was sponsored by the Alexander von Humboldt foundation and the
U.S. Dept. of Energy grant DE-FG03-90ER45435-M012.

\end{document}